\documentclass[a4paper]{spie}  

\usepackage[]{graphicx}

\title{Ultra-strong laser pulses: streak-camera for gamma-rays via pair production and quantum radiative reaction}

\author{K. Z. Hatsagortsyan\supit{a}, A. Ipp\supit{b}, J. Evers\supit{a}, A. Di Piazza\supit{a} and C. H. Keitel\supit{a}
\skiplinehalf
\supit{a}Max-Planck-Institut f\"ur Kernphysik, Saupfercheckweg 1, D-69117 Heidelberg, Germany; \\
\supit{b} Institut f\"ur Theoretische Physik, Technische Universit\"at Wien, 1040 Vienna, Austria
}

\authorinfo{Send correspondence to K. Z. H., e-mail: k.hatsagortsyan@mpi-k.de}

 \date{\today}

\begin{document} 
  \maketitle 

\begin{abstract}
We show that a strong laser pulse combined with a strong x-ray pulse can be employed in a detection scheme for characterizing high-energy $\gamma$-ray pulses down to the zeptosecond timescale. 
The scheme employs streak imaging technique built upon the high-energy process of electron-positron pair production in vacuum through the collision of a test pulse with intense laser pulses. The role of quantum radiation reaction in multiphoton Compton scattering process and limitations imposed by it on the detection scheme are examined. 
\end{abstract}


\section{INTRODUCTION}
\label{sec:intro}  

In recent years attoscience has emerged \cite{Krausz2009}  allowing time-resolved studies of atomic processes. State-of-the-art light sources are capable of producing pulse trains  \cite{Paul2001,Sansone} as well as  single attosecond extreme ultraviolet (XUV) pulses~\cite{Hentschel:2001,Goulielmakis}. For time-resolving the intra-nuclear dynamics \cite{Ledingham,PHOBOS}, $\gamma$-rays with zeptosecond duration and with photon energies larger than MeV are required. There are suggestions to produce zeptosecond pulses of high-energy photons by employing the relativistic oscillating mirror of an overdense plasma surface in a strong laser field \cite{gordienko04,Nomura}, the laser light reflection by a relativistically moving electron density modulation (flying mirror) in a plasma wake of an intense laser pulse \cite{Bulanov,Kando} or via nonlinear Thomson/Compton backscattering \cite{Hartemann,lan:066501,Kim}. Double pulses of yoctosecond duration of GeV photon energy could be created in non-central heavy ion collisions~\cite{Ipp:2009ja}.

For the characterization of short light pulses, a variety of methods is employed. Autocorrelation schemes use the test pulse and its time-shifted replica (FROG~\cite{FROG,Mairesse2005}) or the time- and frequency-shifted replica (SPIDER~\cite{Iaconis1998,Quere}),  while cross-correlation  schemes are based on the correlation between the test XUV pulse and a femtosecond infrared laser pulse. The latter can be weak, inducing few photon effects (RABBITT~\cite{Paul2001}) or strong, yielding attosecond streak imaging \cite{Drescher2001,Itatani2002,Kitzler2002}. Streak imaging~\cite{Drescher2001} is a powerful yet conceptually simple method, in which a short test pulse (TP) and a streaking pulse (SP) co-propagate. A nonlinear mechanism  converts photons from the TP to electrons in the presence of the SP. The final momentum distribution of photo electrons depends on the phase of the SP at the electron emission moment and hence provides information on the duration and the chirp of the short pulse. The efficiency of the streaking is directly related to the conversion mechanism of photons to electrons that depends on the photon energy. 	
The atomic photoionization or Compton ionization  can be used as the conversion mechanism of streaking in the case of XUV or hard x-ray photons. For short pulses of $\gamma$-rays a new conversion mechanism is required to realize streaking  in the sub-attosecond and super-MeV regime.

In this paper, we discuss a scheme for the
characterization of short $\gamma$-ray pulses of GeV energy
photons down to the zeptosecond scale. The  concept of the streaking at high energies with electrons and
positrons (SHEEP)  is based on the  strong field
electron-positron pair production from vacuum by a
$\gamma$-photon of the TP assisted by a
counter-propagating intense laser pulse (IP) \cite{Letter}. A SP co-propagating
with the TP then modifies the kinetics of the created particles
depending on the relative phase of the point of creation within
the SP. Thus, by measuring simultaneously the momentum and energy
of the final electrons and positrons, the TP length and, in
principle, the TP shape can be reconstructed. We  discuss effects limiting the resolution and the detectable
photon energies.   

\section{The SHEEP concept }
\label{concept}

The $\gamma$-photons of the TP under characterization collide with a counter-propagating infrared IP of  linear polarization and are converted into electron-positron pairs. The SP  co-propagates with the $\gamma$-ray pulse and is linearly polarized. For simple streaking dynamics, the polarization of the SP and IP are chosen to be perpendicular, with IP polarized along the $x$- and the SP along the $y$-axis. 
The most important requirement for SHEEP is that a sufficient number of
electron-positron pairs is created by the laser fields. The strong
field pair production process is governed by two relativistic
invariant parameters $\xi=e \sqrt{A_{\mu}A^{\mu}}/m$ and
$\chi=e\sqrt{(F_{\mu\nu}k_t^{\nu})^2}/m^3$~\cite{Ritus}, where
$A_{\mu}$ and $F_{\mu \nu}$ are the vector potential and the field
tensor of the laser fields, respectively, and $e=-|e|$ and $m$ are the
 charge and the mass of the electron.
In the chosen geometry $\chi=(k_ik_t)\xi_i/m^2=2\omega_i\omega_t\xi_i/m^2$ 
and $\xi^2=\xi^2_i+\xi^2_s$, with the 4-momenta $k_i,k_t$ and frequencies $\omega_i,\omega_t$ of IP and TP, respectively and the field parameters $\xi_i$, $\xi_s$ of the IP and SP, respectively. In our case, $\chi$ depends only on
the field of the counter-propagating IP via $\xi_i$ because the role of the
intense laser field in the pair production process by a $\gamma$-photon
is the compensation of the momentum of the $\gamma$-photon which the
co-propagating SP photons cannot fulfill.  The pair production probability in a counter-propagating laser field is exponentially damped when $\chi \ll 1$: $w_{e^+e^-}\propto \chi^{3/2}\exp(-8/3\chi)$, while $w_{e^+e^-}\propto \chi^{2/3} $ at $\chi \gg 1$ \cite{Ritus}.
The exponential suppression of the pair production probability is avoided if
$ \chi> 8/3$. The condition $\chi \sim 1$ is equivalent to $n_{i0}\sim n_c$, where $n_{i0}$ is the number of absorbed laser photons at the threshold of the process, $n_c\sim \xi_{i}^{3}$ is the characteristic number of photons involved in the strong field multiphoton processes. An expression for the number of absorbed laser photons can be derived from the energy and momentum conservation at the threshold of pair production: $n_{i0}=m_{*}^{2}/(\omega_{i}\omega_{t})$. In fact, in the center-of-mass frame of the produced particles the energy-momentum conservation yields $2\tilde{\gamma}_{cm}n_{i0}\omega_i=\omega_t/2\tilde{\gamma}_{cm}=m_*$, where $\tilde{\gamma}_{cm}$ is the Lorentz-factor of the center-of-mass frame. At $\chi \approx 8/3$, the number of produced pairs can be estimated \cite{Ritus} 
\begin{equation}
\label{Nee} N_{e^+e^-}\sim 10^{-2}\alpha
(m^2/\omega_t)N_t\tau_i, 
\end{equation}
where $\alpha$ is the fine structure constant, $N_t$ the number of photons in the TP, and $\tau_i$ the IP duration ($\hbar=c=1$ units are used throughout the paper).

\section{Achievable resolution}
\label{concept}

Electron and positron are created in a certain phase of the SP. During the further motion of the electron (positron) in field of IP and SP, the signature of the initial phase of the SP in the electron (positron) energy exchange with the laser fields will be maintained if the electron momentum is far from the resonance condition corresponding to the stimulated Compton process driven by the SP and IP: $\omega^{\prime}_i \gg \omega^{\prime}_s$, where $\omega^{\prime}_i=2\gamma_R\omega_i$, $\omega^{\prime}_s=\omega_s/2\gamma_R$ are the Doppler-shifted frequencies of the IP and SP in the electron rest frame, respectively. In the Lab frame this condition reads  $2 \omega_i  \omega_t^2  \gg  \omega_s \xi_i^2/m^2$ \cite{Letter}.
The electrons (positrons) created at different initial phases of the SP gain different amounts of energy during the motion in the laser fields. The larger the energy gain, the higher the streaking resolution. In particular, the SHEEP will work if the energy difference $\Delta\mathcal{E}$ due to streaking  for any two electrons created in TP  exceed the energy uncertainty of the TP, $\Delta\mathcal{E}\gg 1/\tau_{t}$, as well as the bandwidth $\Delta\omega_{t}$ of the $\gamma$-ray beam  $\Delta\mathcal{E}\gg \Delta\omega_{t}$. In this section, we calculate  the electron (positron) energy gain $\Delta\mathcal{E}$ during the motion in the superposition of the IP and SP using relativistic classical equations of motion for estimation of the resolution limits.

The transversal components of the electron momentum with respect to the laser propagation direction $z$ are immediately derived from the canonical momentum conservation,
\begin{equation}
\label{pxpypz}
  p_x=q_x-eA_i(\eta)\,,\quad
  p_y=q_y-eA_s(\zeta)+eA_s(\zeta_0)\,,
\end{equation}
where $\eta=\omega_i(t-z)$ and $\zeta=\omega_s(t+z)$. The initial conditions are determined as follows. Due to the off-resonance condition $\omega^{\prime}_i \gg \omega^{\prime}_s$, the variation of the IP phase $\eta$ is much larger than  $2\pi$ during the electron (positron) creation time, while the phase $\zeta$ is almost constant.   Accordingly, the electron
is born at a $\zeta=\zeta_0$ with an average (over the phase $\eta$) momentum  $\textbf{q}=(q_x,q_y,q_z)$: $\langle \textbf{p} \rangle_{\eta}|_{\zeta=\zeta_0}=\textbf{q}$. 
The equations for the longitudinal momentum $p_z$ and the energy ${\cal E}$ read:
\begin{eqnarray}
    \frac{dp_z}{dt}&=&e\beta_xE_i(\eta)-e\beta_yE_s(\zeta)\,, \\
    \frac{d {\cal E}}{dt}&=&e\beta_xE_i(\eta)+e\beta_yE_s(\zeta), 
\end{eqnarray}
where $\beta$ is the electron velocity in units of the speed of light. From the latter, the quantities $\Lambda\equiv {\cal E}-p_z$ and $\Pi\equiv {\cal E}+p_z$ obey the following equations
\begin{eqnarray}
    \frac{d\Lambda}{dt}&=&2e\frac{p_yE_s(\zeta)}{{\cal E}}\,, \\
    \frac{d\Pi}{dt}&=&2e\frac{p_xE_i(\eta)}{{\cal E}}\,.
\end{eqnarray}
We introduce independent variables
$\eta,\zeta$: $d/dt=\omega_i(1-\beta_z)\partial /\partial\eta+(1+\beta_z)\omega_s\partial /\partial\zeta$ which describe two time scales in the electron dynamics. Due to the off-resonance condition,  a small parameter
$\epsilon=\omega^{\prime}_s/\omega^{\prime}_i=(\omega_s/\omega_i)(\Pi/\Lambda)\ll
1$ arises in the equations of motion 
\begin{eqnarray}\label{lambda}
      \frac{\partial\Lambda}{\partial\eta}+\epsilon\frac{\partial\Lambda}{\partial\zeta}
      &=& -2\frac{\left[q_y+eA_s(\zeta_0)-eA_s(\zeta)\right]eA_s^{\prime}(\zeta)}{\Pi}\epsilon\,,\\
     \frac{\partial\Pi}{\partial\eta}+\epsilon\frac{\partial\Pi}{\partial\zeta}
      &=&-2\frac{[q_x-eA_i(\eta)]eA_i^{\prime}(\eta)}{\Lambda}\,.\label{pi}
\end{eqnarray}
We solve Eqs.~(\ref{lambda})-(\ref{pi}) by the perturbation theory with respect to $\epsilon$: $\Lambda=\Lambda^{(0)}+\Lambda^{(1)}$ and $\Pi=\Pi^{(0)}+\Pi^{(1)}$, where $\Lambda^{(n)},\Pi^{(n)}\sim \epsilon^n$. The following initial conditions are used upon switching off SP ($A_s(\zeta)=A_s(\zeta_0)$):  
\begin{eqnarray} 
\label{Pi0}
\Lambda |_{\xi_s=0}&=& q_0-q_z\\
 \Pi|_{\xi_s=0}&=&\frac{q_{\bot}^2+m^2-2q_xeA_i(\eta)+e^2A_i^2(\eta)}{q_0-q_z}.
\end{eqnarray}
While the initial conditions upon switching off IP ($A_i \rightarrow 0$) are 
\begin{eqnarray} 
 \Pi|_{\xi_i=0}&=& q_0+q_{z} \\ \label{Lambda0}
\Lambda|_{\xi_i=0}&=&\frac{q_{\bot}^2+m^2+2q_ye[A_s(\eta)-A_s(\eta_0)]+e^2[A_s(\eta)-A_s(\eta_0)]^2}{q_0+q_z}.
\end{eqnarray}
Taking into account that 
\begin{eqnarray} 
\Lambda \Pi=\tilde{q}^2_{\bot}+m^2-2q_xeA_i(\eta)+e^2A_i^2(\eta)-2\tilde{q}_yeA_s(\zeta)+e^2A_s^2(\zeta),
\end{eqnarray}
with $\tilde{q}^2_{\bot}=q_x^2+\tilde{q}_y^2$ and $\tilde{q}_y\equiv q_y+A_s(\zeta_0)$, the Eqs.(\ref{lambda})-(\ref{pi}) in $\epsilon^0$-th order read:
\begin{eqnarray}
\label{Lambda1}
      \frac{\partial\Lambda^{(0)}}{\partial\eta}&=&0\,,\\
\label{Pi1}
     \frac{\partial\Pi^{(0)}}{\partial\eta}
      &=&-\frac{2[q_x-eA_i(\eta)]eA_i^{\prime}(\eta)\Pi^{(0)}}{\tilde{q}^2_{\bot}+m^2-2q_xeA_i(\eta)+e^2A_i^2(\eta)-2\tilde{q}_yeA_s(\zeta)+e^2A_s^2(\zeta)}\,.\label{pi0}
\end{eqnarray}
From Eqs. (\ref{Lambda1}-\ref{Pi1}), 
\begin{eqnarray}
\Lambda^{(0)}&=&f(\zeta)\\   \Pi^{(0)}&=&[\tilde{q}^2_{\bot}+m^2-2q_xeA_i(\eta)+e^2A_i^2(\eta)-2\tilde{q}_yeA_s(\zeta)+e^2A_s^2(\zeta)]g(\zeta),
\end{eqnarray}
with arbitrary functions $f(\zeta)$ and $g(\zeta)$. The latter are determined from the initial conditions Eqs. (\ref{Pi0}-\ref{Lambda0}), yielding:
\begin{eqnarray} 
\Lambda^{(0)} &=&\frac{q_{\bot}^2+m_*^2-2q_ye[A_s(\eta)-A_s(\eta_0)]+e^2[A_s(\eta)-A_s(\eta_0)]^2}{q_0+q_z},\\
\Pi^{(0)} &=&\frac{q_{\bot}^2+m^2}{q_0-q_z}\frac{q_{\bot}^2+m^2-2q_xeA_i(\eta)+e^2A_i^2(\eta)+2q_ye[A_s(\eta)-A_s(\eta_0)]+e^2[A_s(\eta)-A_s(\eta_0)]^2}{q_{\bot}^2+m^2+2q_ye[A_s(\eta)-A_s(\eta_0)]+e^2[A_s(\eta)-A_s(\eta_0)]^2}
\end{eqnarray}
After the interaction with the IP and SP $A_i(\eta),A_s(\zeta)\rightarrow 0$, 
\begin{eqnarray} 
\Lambda^{(0)} &=&\frac{q_{\bot}^2+m_*^2-2q_yeA_s(\eta_0)+e^2A_s^2(\eta_0)}{q_0+q_z},\\
\Pi^{(0)} &=&\frac{q_{\bot}^2+m^2}{q_0-q_z},
\end{eqnarray}
which determine the electron energy after the interaction 
\begin{equation}
\label{energy}
    {\cal E}=q_0-\frac{m^2\xi_i^2}{4(q_0-q_z)}+\frac{q_yeA_s(\zeta_0)}{q_0+q_z}
    +\frac{e^2A_s^2(\zeta_0)}{2(q_0+q_z)}\,.
\end{equation}
The electron and the positron are produced not only at the threshold
with zero momentum in the center-of-mass frame but also
above-threshold. The number of absorbed IP photons at the threshold is $n_{i0}=m_*^2/\omega_t\omega_i$.
The width of variation of the absorbed laser photons ($n_i$) from the threshold value ($n_{i0}$)
is of order $\delta n_i\sim n_{i0}$~\cite{Ritus}. Absorbing $n_i$ photons from the laser field, the particles in the center-of-mass frame are born with an energy  ${\cal E}_{cm}=\sqrt{n_i\omega_i\omega_t}$ and  with the polar emission angles  $\theta,\phi$ for the positron. 
In fact, ${\cal E}_{cm}=\sqrt{n_i\omega_i\omega_t}$ follows from
  the energy-momentum conservation in the center-of-mass frame:
  $2\tilde{\gamma}_{cm}n_{i}\omega_i=\omega_t/2\tilde{\gamma}_{cm}={\cal
  E}_{cm}$.
The momenta  and energy of the particles in the lab frame  then are $p_{x0}^{\pm}=\pm\sqrt{\omega_tn_i\omega_i}\delta \sin\theta\cos\phi$, $p_{y0}^{\pm}=\pm\sqrt{\omega_tn_i\omega_i}\delta \sin\theta\sin\phi$ and ${\cal E}_{0}^{\pm}=(\omega_t+n_i \omega_i)(1\mp\beta_n\delta\cos\theta)/2$. Here, $\pm$ indices correspond to the  positron and electron, respectively, and $\delta\equiv \sqrt{\delta n_i/n_{i}}< 1/\sqrt{2}$, and $\beta_n\equiv  (\omega_t-n_i \omega_i)/(\omega_t+n_i \omega_i)\approx 1$. After the interaction with the laser fields ($A_i(\eta),A_s(\zeta) \rightarrow 0$), the momenta and energy of the particles are given by
\begin{eqnarray}
 p_{x}^{\pm}&=&\pm m_*\delta\sin\theta\cos\phi/\sqrt{1-\delta^2}\,, \\
p_{y}^{\pm}&=&\pm m_*\delta\sin\theta\sin\phi/\sqrt{1-\delta^2} \mp eA_s(\zeta_0)\,, \\
{\cal E}_{0}^{\pm}&\approx & \frac{\omega_t}{2}\left[ 1\mp\beta_n\delta \cos \theta +\frac{2\delta\sin\theta\sin\phi\sqrt{1-\delta^2}eA_s(\zeta_0)}{(1\pm\delta\cos\theta)m_*}
-\frac{m^2\xi_i^2}{2\omega_t^2(1\mp\delta\cos\theta)}+\frac{e^2A_s^2(\zeta_0)(1-\delta^2)}{(1\pm\delta\cos\theta)m_*^2}\right].
\end{eqnarray}
Since the values $\{\theta,\phi,\delta,\zeta_0\}$ can be deduced from the measured $\{p_x,p_y,{\cal E}^+,{\cal E}^-\}$, the coincidence measurement of the electron and positron momenta after the interaction  provides  information on the pair production phase $\zeta_0$ in the SP. The  energy difference $\Delta\mathcal{E}$ of two electrons created at two different $\zeta_{1}$ and $\zeta_{2}$ can be derived using the following expressions: $A_s(\zeta_{2})-A_s(\zeta_{1})\approx-E_s(\zeta_{0})(\zeta_{2}-\zeta_{1})/\omega_s$, $A_s^2(\zeta_{2})-A_s^2(\zeta_{1})\approx-2A_s (\zeta_{0})E_s(\zeta_{0})(\zeta_{2}-\zeta_{1})/\omega_s$ and $\zeta_{2}-\zeta_{1}=\omega_s\tau_t$:
\begin{equation}
 \Delta\mathcal{E}\sim \omega_{t}\omega_{s}\tau_{t} \max\left\{\frac{\xi_s}{\sqrt{2}\xi_i},\frac{\xi_{s}^{2}}{\xi_i^2}\right\}.
\label{deltaE}
\end{equation}
Using Eq.~(\ref{deltaE}) and assuming $\xi_i\gg \xi_s$, the resolution conditions $\Delta\mathcal{E}\gg 1/\tau_{t},\Delta \omega_t$ become
\begin{eqnarray}
\label{eq:streak}
(\omega_s\tau_t)^2 &\gg& (\omega_s/\omega_t)(\xi_i/\xi_s),\\
\Delta  \omega_t/\omega_t &\ll& \omega_s\tau_t(\xi_s/\xi_i).
\label{eq:streak2}
\end{eqnarray}
At last, basic preconditions for streak imaging should be mentioned that the TP length
$\tau_{t}$ is shorter than half of the SP wavelength
$\lambda_{s}=2\pi /\omega_{s}$, and that the
streaking signal exceeds the noise level~\cite{Krausz2009},
\begin{equation}
\pi N/S \ll \omega_{s}\tau_{t}<\pi\,,
\label{eq:SN}
\end{equation}
where $S/N$ is the signal-to-noise ratio for the laser fields.

\section{Radiation reaction}
\label{RDR}

In this section we discuss the role of radiation reaction in the SHEEP operation. The electron (positron) moving in a strong laser field can radiate via multiphoton Compton scattering which may alter the electron dynamics and disturb the SHEEP operation. Let us estimate the radiation back reaction.
The photon emission probability during the motion on the radiation formation length is of order of $\alpha$. The characteristic energy of the emitted photon in the electron average rest frame (R-frame) is $\omega_c^{\prime}\sim \bar{n}\omega_0^{\prime}$, where $\omega_0^{\prime}$  is the laser frequency in the  R-frame, $\bar{n}\sim\xi^3$ is the effective number of the emitted harmonics \cite{Ritus}. According to \cite{Salamin}, $\omega_0^{\prime}=(\varepsilon_0+p_0)\omega_0/\varepsilon^{\prime}_0$, with the electron initial energy $\varepsilon_0$ and momentum $p_0$ before entering the laser pulse in the Lab frame and  the electron energy in the R-frame $\varepsilon^{\prime}_0\sim m\xi$. As a result, $\omega_c^{\prime}\sim \chi\varepsilon^{\prime}_e$, with the $\chi$ parameter of the electron $\chi=\xi(\varepsilon_0+p_0)\omega_0/m^2$.  Therefore, the radiated energy in the R-frame during the radiation formation time is $\Delta \varepsilon_{rad}^{(f)\prime} \sim \alpha \chi \varepsilon^{\prime}_0$. When the electron moves in a laser field, the phase interval corresponding to the formation time is $\Delta \phi_f \sim 1/\xi$ \cite{Ritus}, the radiated energy during the laser period, i. e. within a phase interval of $\Delta \phi \sim 1$, will be: $\Delta \varepsilon_{rad}^{(T)\prime}\sim \alpha \omega_c\varepsilon^{\prime}_0 \Delta \phi/\Delta \phi_c=\alpha \xi\chi\varepsilon^{\prime}_0$. The radiation reaction will be significant if the energy loss of the electron due to radiation during the motion in one laser period is comparable with the electron initial energy: $\Delta \varepsilon_{rad}^{(T)\prime} > \varepsilon^{\prime}_0$. The latter condition  reads $\alpha \xi \chi> 1$. The probability of a photon emission during a phase interval $\Delta \phi$ is $w\sim \alpha \xi  \Delta \phi$ and during one period $w_T\sim \alpha \xi $. The parameter indicating the role of the emitted photon recoil is $\omega_c^{\prime}/\varepsilon^{\prime}_0\sim \chi$. When $\chi\ll 1$ the photon recoil is negligible and the classical description is valid.
Even though in classical theory the radiation reaction force is a perturbation in the rest frame of the electron, in the Lab-frame it can be the dominant force \cite{Landau2}. Accordingly, a 
radiation dominated regime (RDR) for Thomson scattering can be identified  \cite{Koga_2005,Di_Piazza_2008} when the radiated energy during the driving laser period is of order or exceeds the electron energy. The latter condition determines the RDR parameter:  $R\equiv \alpha \xi \chi > 1 $.

When $\chi> 1$ the quantum effects become conspicuous. 
The effects of radiation reaction in the quantum regime can be classified as a) quantum effects due to close coupling of different channels of photon emission because of rescattering of photons by the radiating electron \cite{Heitler} and b) quantum effects due to the multiple recoils experienced by the electron in emitting photons \cite{Baier_b_1994}. 
In the usual formalism of perturbative QED, the quantum effects a) are compounded with radiative corrections \cite{Heitler_book} and carry out all singularity problems transferred from the classical radiation reaction theory. The effects a) can have influence on the radiation formation length. This influence can be quantified by a parameter $\kappa$ which is the ratio of the electron energy change due to radiation reaction on the radiation formation  length to the electron energy: $\kappa\equiv \Delta \varepsilon_{rad}^{(f)\prime}/\varepsilon^{\prime}_0\sim \alpha\chi$.   
The mentioned effect can be significant $\kappa >1$ mostly  in the region of nonperturbative QED  ($\alpha \chi^{2/3}> 1$) \cite{Ritus}. Moreover, the latter region  is not achievable by laser fields in foreseeable future and we will not touch upon the effects of kind a). In a domain of $\alpha \chi\ll 1$, Compton scattering in the radiation formation length is not disturbed by the radiation reaction effects of type a), while those of kind b) must be taken into account if $\chi> 1$. The quantum recoil in the emission of one photon is incorporated in the standard QED via the energy-momentum conservation delta functions and does not require special consideration. However, the electron can emit successively multiple photons. When $\alpha \xi \gg 1$  multiple photon emission happens during one period  which can take place in experiments with existing strong lasers intensities exceeding $10^{22}$ W/cm$^2$ \cite{Yanovsky_2008}. Then, to consider the quantum effects b), one has to take into account  the change of the electron state in each successive photon emission event which happens in a statistically uncorrelated way. 
Radiation reaction effects in the interaction of an electron and a strong laser field in the quantum regime are investigated in the work \cite{QRDR}. The quantum radiation reaction is identified with the 
multiple photon recoils experienced by the laser-driven electron due to consecutive incoherent photon emissions. In this regime radiation reaction affects multiphoton Compton scattering spectra which can be measurable with already presently available laser systems. In particular, inclusion of the RR in the quantum regime  increases of the spectral yield at low energies,   shifts to lower energies of the maximum of
the spectral yield, and  decreases of the spectral yield at high energies (photon piling).

Returning to the SHEEP problem, we estimate  the probability of a photon emission in the multiphoton Compton process $w_C\sim \alpha \xi_i \mathcal{N}_i$ in the strong field of IP with number of cycles $\mathcal{N}_i$. The photon emission will be negligible when
\begin{equation}
\alpha \xi_i \mathcal{N}_i\ll 1,
 \label{C9}
\end{equation}
In the streaking regime $\chi \sim 1$, therefore, $\alpha \xi \chi \ll 1$, while only in the opposite limit, $\alpha \xi \chi > 1$, the radiation dominated regime of multiphoton Compton scattering is entered. 
This condition can be weakened to $\alpha \xi_i \mathcal{N}_i\sim 1$ by
selectively dropping Compton scattering events, which can be
identified by comparing momenta of the electron and positron after
the interaction.

\section{The SHEEP parameters}
\label{parameters}

There are different possibilities to realize SHEEP. The IP should be a short and relatively
strong laser field with $\xi_i\sim 1-10$, ${\mathcal N}_i=3-30$ as
required from Eq.~(\ref{C9}). The minimal photon energy of the TP
depends on the IP frequency and intensity, given by the condition $\chi \sim 1$. Thus, at an infrared IP
with $\xi_i=10$, corresponding to a laser intensity of
$I_i=10^{20}$ W/cm$^2$, one obtains $\omega_{t\,min}=30$ GeV,
while in the case of an ultraviolet IP with $\xi_i=1$
($\omega_i=1000$ eV, $I_i=10^{24}$ W/cm$^2$), instead
$\omega_{t\,min}=300$ MeV. Three regimes for SHEEP can exist with SP of
different frequency: femtosecond TP with $\omega_s=1$ eV,
attosecond TP with $\omega_s=100$ eV and zeptosecond TP with
$\omega_s=1$ keV. 
If the TP bandwidth is $\Delta \omega_t/\omega_t\sim 0.1$,
$\xi_s/\xi_i> 0.1$ will be required. The required infrared
IP with an intensity of $10^{20}$ W/cm$^2$ is routinely available
in many labs. The intense high-frequency SP/IP with photon energies
in the $0.1-1$ keV range can be produced in the ELI facility via
high-order harmonic generation at plasma surfaces~\cite{ELI}.

\section{Conclusion}
\label{conclusion}

We have presented a detection scheme for the
characterization of short $\gamma$-ray pulses of the GeV photon
energy range. Sub-attosecond time resolution could be achieved in the upcoming
ELI facility using strong XUV pulse via high-order harmonic generation at overdense plasma surface. 
Restricting the intensity of the applied infrared laser field, radiation reaction effects can be avoided.


\bibliography{sheep_spie}   

\begin{thebibliography}{10}

\bibitem{Krausz2009}
Krausz, F. and Ivanov, M., ``Attosecond physics,'' {\em Rev. Mod. Phys.}~{\bf
  81},  163--234 (2009).

\bibitem{Paul2001}
{Paul, \textit{et al.}}, ``Observation of a train of attosecond pulses from
  high harmonic generation,'' {\em Science}~{\bf 292},  1689 (2001).

\bibitem{Sansone}
{Sansone, G., \textit{et al.}}, ``Isolated single-cycle attosecond pulses,''
  {\em Science}~{\bf 314},  443--446 (2006).

\bibitem{Hentschel:2001}
{Hentschel, \textit{et al.}}, ``Attosecond metrology,'' {\em Nature}~{\bf 414},
   509--513 (2001).

\bibitem{Goulielmakis}
{Goulielmakis E., \textit{et al.}}, ``Single-cycle nonlinear optics,'' {\em
  Science}~{\bf 320},  1614--1617 (2008).

\bibitem{Ledingham}
Ledingham, K. W.~D., P.~M. and Singhal, R.~P., ``Applications for nuclear
  phenomena generated by ultra-intense lasers,'' {\em Science}~{\bf 300},  1107
  (2003).

\bibitem{PHOBOS}
{Back B. B. \textit{et al.}}, ``Significance of the fragmentation region in
  ultrarelativistic heavy-ion collisions,'' {\em Phys. Rev. Lett.}~{\bf 85},
  823--826 (2000).

\bibitem{gordienko04}
{Gordienko, S., Pukhov, A., Shorokhov, O., and Baeva, T.}, ``Relativistic
  {Doppler} effect: {Universal} spectra and zeptosecond pulses,'' {\em Phys.
  Rev. Lett.}~{\bf 93},  115002 (2004).

\bibitem{Nomura}
{Nomura Y. \textit{et al.}}, ``Attosecond phase locking of harmonics emitted
  from laser-produced plasmas,'' {\em Nature Phys.}~{\bf 5},  124--128 (2009).

\bibitem{Bulanov}
{Bulanov, S. V., Esirkepov, T. Zh., and Tajima, T.}, ``Light intensification
  towards the schwinger limit,'' {\em Phys. Rev. Lett.}~{\bf 91},  085001
  (2003).

\bibitem{Kando}
{Kando, \textit{et al.}}, ``Enhancement of photon number reflected by the
  relativistic flying mirror,'' {\em Phys. Rev. Lett.}~{\bf 103},  235003
  (2009).

\bibitem{Hartemann}
{Hartemann F. V. \textit{et al.}}, ``High-energy scaling of compton scattering
  light sources,'' {\em Phys. Rev. ST AB}~{\bf 8},  100702 (2005).

\bibitem{lan:066501}
{Lan P. \textit{et al.}}, ``Attosecond and zeptosecond x-ray pulses via
  nonlinear thomson backscattering,'' {\em Phys. Rev. E}~{\bf 72},  066501
  (2005).

\bibitem{Kim}
{Kim D. \textit{et al.}}, ``Attosecond kev x-ray pulses driven by thomson
  scattering in a tight focus regime,'' {\em New J. Phys.}~{\bf 11},  063050
  (2009).

\bibitem{Ipp:2009ja}
Ipp, A., Keitel, C.~H., and Evers, J., ``Yoctosecond photon pulses from
  quark-gluon plasmas,'' {\em Phys. Rev. Lett.}~{\bf 103},  152301 (2009).

\bibitem{FROG}
Kane, D.~J. and Trebino, R., ``Single-shot measurement of the intensity and
  phase of an arbitrary ultrashort pulse by using frequency-resolved optical
  gating,'' {\em Opt. Lett.}~{\bf 18},  823--825 (1993).

\bibitem{Mairesse2005}
Mairesse, Y. and Qu\'er\'e, F., ``Frequency-resolved optical gating for
  complete reconstruction of attosecond bursts,'' {\em Phys. Rev. A}~{\bf 71},
  011401 (Jan 2005).

\bibitem{Iaconis1998}
Iaconis, C. and Walmsley, I., ``Spectral phase interferometry for direct
  electric-field reconstruction of ultrashort optical pulses,'' {\em Opt.
  Lett.}~{\bf 23},  792--794 (1998).

\bibitem{Quere}
{Qu\'er\'e, F., Itatani, J., Yudin, G. L., and Corkum, P. B.}, ``Attosecond
  spectral shearing interferometry,'' {\em Phys. Rev. Lett.}~{\bf 90},  073902
  (2003).

\bibitem{Drescher2001}
{Drescher M. \textit{et al.}}, ``X-ray pulses approaching the attosecond
  frontier,'' {\em Science}~{\bf 291},  1923--1927 (2001).

\bibitem{Itatani2002}
{Itatani J. \textit{et al.}}, ``Attosecond streak camera,'' {\em Phys. Rev.
  Lett.}~{\bf 88},  173903 (2002).

\bibitem{Kitzler2002}
{Kitzler M. \textit{et al.}}, ``Quantum theory of attosecond xuv pulse
  measurement by laser dressed photoionization,'' {\em Phys. Rev. Lett.}~{\bf
  88},  173904 (2002).

\bibitem{Letter}
{Ipp A., \textit{et al.}}, ``Streaking at high energies with electrons and
  positrons,''  to be published.

\bibitem{Ritus}
Ritus, V.~I., ``Quantum effects of the interaction of elementary particles with
  an intense electromagnetic field,'' {\em J. Sov. Laser Res.}~{\bf 42},
  497--615 (1985).

\bibitem{Salamin}
Salamin, Y.~I. and Faisal, F. H.~M., ``Harmonic generation by superintense
  light scattering from relativistic electrons,'' {\em Phys. Rev. A}~{\bf 54},
  4383--4395 (1998).

\bibitem{Landau2}
Landau~L.D., Lifshits, E.~M.,  [{\em The Classical Theory of
  Fields}{\nolinebreak\hspace{0.1em}]}, Elsevier, Oxford (1975).

\bibitem{Koga_2005}
{Koga, J. \textit{et al.} } {\em Phys. Plasmas}~{\bf 12},  093106 (2005).

\bibitem{Di_Piazza_2008}
Di~Piazza, A. {\em Lett. Math. Phys.}~{\bf 83},  305 (2008).

\bibitem{Heitler}
Heitler, W. {\em Proc. Camb. Phil. Soc.}~{\bf 37},  291 (1941).

\bibitem{Baier_b_1994}
Baier, V.~N., K. V.~M. and Strakhovenko, V.~M.,  [{\em Electromagnetic
  Processes at High Energies in Oriented Single
  Crystals}{\nolinebreak\hspace{0.1em}]}, World Scientific, Singapore (1994).

\bibitem{Heitler_book}
Heitler, W.,  [{\em The quantum theory of
  radiation}{\nolinebreak\hspace{0.1em}]}, Clarendon Press, Oxford (1954).

\bibitem{Yanovsky_2008}
{Yanovsky, V. \textit{et al. }} {\em Opt. Express}~{\bf 16},  2109 (2008).

\bibitem{QRDR}
{Di Piazza, A., Hatsagortsyan, K. Z., Keitel, C. H.}, ``Quantum radiation
  reaction effects in multiphoton compton scattering,'' {\em Phys. Rev.
  Lett.}~{\bf 105},  220403 (2010).

\bibitem{ELI}
{F. Amiranoff \textit{et al.}} Proposal for a European Extreme Light
  Infrastructure (ELI).

\end{thebibliography}
\bibliographystyle{spiebib}   

\end{document}